\documentclass[12pt]{article}

\usepackage{graphicx}

\usepackage{color} 
\usepackage{slashed}
\usepackage{amsmath}
\usepackage[pdfstartview=FitH,colorlinks=true,linkcolor=black,anchorcolor=black,citecolor=black,urlcolor=black]{hyperref}
\usepackage{amsmath,amssymb,titling,authblk}
\usepackage{slashed}
\usepackage{amsmath,bm}
\usepackage{amscd}
\usepackage[normalem]{ulem}
\usepackage{appendix}
\usepackage{bbold}

\usepackage[utf8]{inputenc}
\usepackage{slashed}
\usepackage[top=2.3cm,right=2.3cm,left=2.3cm,bottom=2.3cm]{geometry}

\definecolor{verde}{cmyk}{.83,.21,1,.08}
\definecolor{darkorchid}{rgb}{0.6, 0.2, 0.8}
\definecolor{darkgreen}{rgb}{0,.5,0}

\def\({\left(}
\def\){\right)}
\def\[{\left[}
\def\]{\right]}

\newcommand{\be}{\begin{equation}}
\newcommand{\ee}{\end{equation}}
\newcommand{\bea}{\begin{eqnarray}}
\newcommand{\eea}{\end{eqnarray}}

\begin{document}

\title{\Large\bf  Effects of wave propagation in canonical Poisson gauge theory under an external magnetic field}

\author[1]{O. Abla}
\author[2]{M. J. Neves}
\affil[ ]{}
\affil[1]{\textit{\footnotesize CCNH - Universidade Federal do ABC, 09210-580, Santo Andr\'e, SP, 
Brazil. }}
\affil[2]{\textit{\footnotesize  Departamento de F\'isica, Universidade Federal Rural do Rio de Janeiro, 23890-971, Serop\'edica, RJ, Brazil.}}
\affil[ ]{}
\affil[ ]{\footnotesize e-mail: \texttt{olavo.abla@ufabc.edu.br, mariojr@ufrrj.br}}

\maketitle

\begin{abstract}
The non-commutative electrodynamics based on the canonical Poisson gauge theory is studied in this paper.
For a pure spatial non-commutativity, we investigate the plane wave solutions in the presence of a constant
and uniform magnetic background field for the classical electrodynamics in canonical Poisson gauge theory. We obtain the
properties of the medium ruled by the permittivity and the permeability tensors in terms of the non-commutative parameter, with the
electrodynamics equations in the momentum space. Using the plane wave solutions mentioned, the dispersion relations
are modified by the magnetic background, and the correspondent group velocity is affected by the spatial non-commutative parameter.
We construct the energy-momentum tensor and discuss the conserved components of this tensor in the spatial non-commutative case.
The birefringence phenomenon is showed through the modified dispersion relations, that depends directly on the non-commutative corrections
and also on the magnetic background field. Using the bound of the polarized vacuum with laser (PVLAS) experiment for the vacuum magnetic
birefringence, we estimate a theoretical value for the spatial non-commutative parameter.
\end{abstract}
\section{Introduction}
In the $1930$'s, Matvei Bronstein introduced a prototype to investigate the quantum effects on weak gravitational fields \cite{Bronstein}. His idea motivated questions about how small the measure of the uncertainty $\Delta x$ for a particle's position can be. In quantum mechanics,
for a test particle, the uncertainty on a measurement of the position must be of the order of wave-length ($\Delta x\sim\lambda=\hbar c/E$), which means that for a higher precision, higher energy is required. Nevertheless, the general relativity states that each mass or energy creates the curvature of the space-time, which is proportional to the Schwarzschild radius $(r_S)$, with $r_S \sim (E/M_{P}) \, \ell_{P}$, where $\ell_P=\sqrt{\hbar G/c^3}=1.6\times 10^{-35}$ m is the Planck length, and $M_{P}=\sqrt{\hbar\,c/G}=2.17 \times 10^{-8}$ kg is the Planck mass \cite{PhysRep}. Thereby, we may postulate that space-time becomes non-local at Planck scale, {\it i.e.}, $\Delta x \,\Delta r_S \geq \ell_P^2$, turning the space-time into a \textit{non-commutative} (NC) manifold \cite{Doplicher}.
The NC structure for space-time coordinates was introduced by Snyder \cite{snyder47} as an attempt to deal with the divergences that emerged on quantum electrodynamics (QED). However, the success of the renormalization procedure on predicting numerical observable of quantum field theory made the NC formalism ignored for many years \cite{SzaboReview}. After decades in the ostracism, the NC approach was reconsidered, specially when it emerged in open string quantization \cite{seibergwitten99}. The NC geometry plays a role in the quantum dynamics of open strings in Neveu-Schwartz two-form $B$-field, which is equivalent to a constant and uniform magnetic background field (MBF) on classical field theories on D-branes. The open string interaction can be controlled by the \textit{Weyl-Groenewold-Moyal} star product \cite{Groenewold,Moyal}. This suggests a NC gauge theory, such as Yang-Mills coupled to scalar fields \cite{SzaboReview}. For a non-constant $B$-field, the theory is not trivial because the $D$-brane needs to live on a curved background and the full machinery of the Kontsevich's quantization has to be implemented \cite{Kontsevich} (see \cite{KV2008} for an alternative approach to the construction of star product).
Although the physics in NC formalism has been highly studied over the last twenty years \cite{Nekrasov}-\cite{Last}, some aspects on the construction of NC gauge theories were not fully understood, mainly when the NC parameter is position-dependent and non-associative \cite{gaugeinvariance}-\cite{KS17}. Some mathematical frameworks such as the correspondence between NC gauge orbits induced by classical gauge orbits, known as the \textit{Seiberg-Witten map} \cite{seibergwitten99,KupAlex}, and the use of higher algebraic structures($L_{\infty}$-algebras) \cite{BBKL}-\cite{KurkovVitale22} were developed to try to solve this puzzle.
To understand the full picture of the NC gauge theories, Kupriyanov et al \cite{Kupriyanov:2020sgx}-\cite{Kup25} have recently proposed the semi-classical
formalism known as {\it Poisson gauge theory} (PGT) \cite{Kup33}. For a latest review on this approach, see the ref. \cite{PhysRep}. Motivated by this new approach on NC gauge theories, we construct the semi-classical limit of the canonical NC electrodynamics based on \cite{Rivelles} to contribute with new ideas using the non-linear electrodynamics approach \cite{Neves2}-\cite{Neves3}. Thereby, we study the classical solutions such as the plane wave (PW) solutions in the presence of a MBF, and the conserved components of the energy momentum tensor (EMT) for the spatial NC case.
The plan of this letter is as follows: In the section 2, we review the structure of the PGT, obtaining the Maxwell-Poisson electrodynamics. In the section 3, we obtain the free wave equation for the four-potential modified by the NC parameter at first order.
Using the PW solutions summed to a MBF, the spatial NC is interpreted as a material medium, in which the permittivity and permeability tensors are written in terms of the NC parameter. Thus, the results for the dispersion relations (DRs) and the group velocity are discussed.
The section 4 is dedicated to the conservation laws for the spatial NC through the symmetric EMT  \cite{emtensor}. The energy density and the Poynting vector are obtained in the linear approximation of the NC parameter and we investigate them using the DRs from the section 3. In the section 5, we show the birefringence phenomenon, which was not obtained on \cite{Rivelles}. Using the bound of the polarized vacuum laser (PVLAS) experiment for the vacuum magnetic birefringence \cite{25years}, we estimate a theoretical value for the spatial NC parameter \cite{Carroll}. Finally, we appoint some conclusions and perspectives for forthcoming projects in the last section.
\section{Maxwell-Poisson equations}
We begin this section with some basic mathematical preliminaries. Let $M$ be a manifold which represents a $n$-dimensional space-time, with local coordinates $x=(x^a)$ ($a=0,1,2,...,n-1$), in which $f=f(x)$ and $g=g(x)$ are two functions $f,g\in C^{\infty}(M)$. The Kontsevich star product of these smooth functions on $M$, which encodes the NC geometry of the space-time, is given by \cite{Kontsevich}
\begin{equation}
	f \star' g:=f \, g+\frac{i\hbar}{2} \, \{\,f\,,\,g\,\}_P \, + \, \mathcal{O}(\hbar^2) \; ,
\end{equation}
with the Poisson bracket defined by
\begin{equation}\label{PB}
	\{\,f\,,\,g\,\}_P=\Theta^{ab}(x) \, \partial_af \, \partial_bg \; .
\end{equation}
The classical limit of the Kontsevich star product, when $\hbar\rightarrow0$, gives directly the standard point-wise product. Taking
the limit $\hbar \rightarrow 0$ in the star commutator of the functions $f$ and $g$,
\begin{equation}
	\{\, f\, , \, g \, \}_P=\lim_{\hbar\to0} (i\hbar)^{-1} \, [ \, f \, , \, g \, ]_{\star'} \; ,
\end{equation}
it yields the \textit{semi-classical} limit. In fact, this approximation assumes that the fields vary slowly, and the higher derivatives terms in the star commutator are negligible compared to the corresponding Poisson bracket, besides it is given in all orders in $\Theta^{ab}$ \cite{Kupriyanov:2020sgx}-\cite{Kup25}. The Kupriyanov's construction of PGT \cite{Kup33} states that, to know the full NC picture, it is reasonable to construct a NC gauge theory using the semi-classical limit, since it satisfies two main properties:
\begin{enumerate}
	\item Gauge invariance ;
	\item Correct commutative limit (usual $U(1)$ gauge theory).
\end{enumerate}
Both statements suggest that if one works with a deformed (nonabelian) $U_{\star'}(1)$ gauge theory, the Poisson gauge algebra is set by $U_P(1)$. In fact, the Moyal star product \cite{Groenewold,Moyal} is the \textit{canonical} version of the Kontsevich product \cite{Kontsevich}
\begin{equation}
	\left. f(x)\star g(x)=\exp\left(\frac{i\hbar}{2} \, \theta^{ab} \, \partial_a^x \, \partial_b^y\right)f(x) \, g(y) \right|_{y\rightarrow x} \; ,
\end{equation}
in which by construction it means that the symplectic two-form $\theta^{ab}$ is the constant version of the Poisson bi-vector $\Theta^{ab}$, with the semi-classical limit of the Moyal commutator equal to the canonical Poisson bracket, {\it i.e.},
\begin{equation}
	(i\hbar)^{-1}\left[ \, f \, , \, g \, \right]_{\star}\approx \left\{ \, f\, , \, g \, \right\}=\theta^{ab} \, \partial_af \, \partial_bg \; ,
\end{equation}
where the gauge algebra that governs the theory is the $U_{\star}(1)$, and then the canonical version of the gauge symmetry
is $U_P^{can}(1)$. The other Poisson brackets involving the variables $x^{a}$-position and $p_{a}$-momentum are as
the usual case \footnote{We use the metric convention $\eta^{ab}=\mbox{diag}\left(+1,-1,-1,-1\right)$.}
\begin{equation}
	\{ \, x^a\, , \, p_b \, \} \,=\,\delta^{a}_{\;\;\,b}
	\hspace{0.3cm} \mbox{and} \hspace{0.3cm}
	\{ \, p_a \, , \, p_b \, \}\,=0 \; .
\end{equation}
We will call the correspondent field equations on the semi-classical limit of full NC electrodynamics as the {\it Maxwell-Poisson} equations \cite{Kup25}.
Using the previous statements, the deformed $U_P^{can}(1)$ gauge group defines the gauge variation
\begin{equation}
	\delta_fA_a:=\partial_af+\left\{ \, A_a \, , \, f \, \right\}=\left\{ \, f \, , \, p_a-A_a \, \right\} \, ,
	\label{a3}
\end{equation}
which gives the following gauge algebra
\begin{equation}
	\left[\,\delta_f \, , \, \delta_g \, \right]=\delta_{\{f \, , \, g\}} \; ,
\end{equation}
and the gauge covariant derivative acting on a $h$-function
\begin{equation}
	{\cal D}_a\,h:=\left\{ \, h \, , \, p_a-A_a \, \right\} \; .
	\label{a7}
\end{equation}
The commutator of two gauge covariant derivatives operators gives
\begin{equation}
	\left[ \, {\cal D}_a \, , \, {\cal D}_b \, \right]\,h=\left\{ \, \mathcal{F}_{ab} \, , \, h \, \right\} \; ,
	\label{a12}
\end{equation}
where $\mathcal{F}_{ab}$ is the field strength tensor
\begin{equation}
	\mathcal{F}_{ab}:=\{p_a-A_a,p_b-A_b\}=
	F_{ab}+\left\{ \, A_a \, , \, A_b \, \right\} \; ,
	\label{strength}
\end{equation}
and $F_{ab}=\partial_{a}A_{b}-\partial_{b}A_{a}$ is the usual Maxwell's strength tensor.
The gauge transformation of $\mathcal{F}_{ab}$ is
\begin{equation}
	\delta_f \mathcal{F}_{ab}=\left\{ \, \mathcal{F}_{ab} \, , \, f \, \right\} \; ,
	\label{a8}
\end{equation}
in which it transforms covariantly under the gauge transformation $\delta_f$.
From these symmetry properties, the covariant lagrangian is constructed as the usual 
\footnote{From this point, we use the natural units in which $\hbar=c=1$, 
	and the electric and magnetic fields have squared-energy dimension. The conversion of tesla (T) to the natural system 
	is $1{\text T} =  6.8 \times 10^{-16} \, \mbox{GeV}^2$. }
\begin{equation}
	{\cal L}_{ED}=-\frac{1}{4} \, \mathcal{F}_{ab}\,\mathcal{F}^{ab}-J^{a}\,A_{a} \; .
	\label{LED}
\end{equation}
which makes the correspondent $4D$ action gauge invariant defined on this coordinate space,
and $J^{a}=(\rho,{\bf J})$ is an external source.
From the action principle applied to the lagrangian (\ref{LED}), the first pair of Maxwell-Poisson equations is
\begin{equation}
	\mathcal{D}_a\mathcal{F}^{ab}=J^{b} \; ,
	\label{FieldEq}
\end{equation}
which recovers the Maxwell's equations with source $J^{a}$ in the commutative limit
\begin{equation}
	\lim_{\theta\rightarrow0}\mathcal{D}_a\mathcal{F}^{ab}=\partial_a F^{ab}=J^{b} \; .
\end{equation}
Using the properties of Poisson brackets, the NC Bianchi identity reads as
\begin{equation}
	\mathcal{D}_a\mathcal{F}_{bc}+\mathcal{D}_b\mathcal{F}_{ca}+\mathcal{D}_c\mathcal{F}_{ab}=0 \; ,
	\label{Bianchi}
\end{equation}
which yields the second pair of Maxwell-Poisson equations. The commutative limit gives the Maxwell equations with no sources
\begin{equation}
	\lim_{\theta\rightarrow0}	\mathcal{D}_a\mathcal{F}_{bc}+\mathcal{D}_b\mathcal{F}_{ca}+\mathcal{D}_c\mathcal{F}_{ab}=	\partial_aF_{bc}+\partial_bF_{ca}+\partial_cF_{ab}=0 \; .
\end{equation}
It is worth to say that this approach is completely equivalent to the first order approximation in $\hbar$ of the field equations on the Moyal picture
\begin{equation}
	\bar{\mathcal{D}}_a\mathcal{F}_M^{ab}=\partial_a\mathcal{F}_M^{ab}-i\left[ \, A_a \, , \, \mathcal{F}_M^{ab} \, \right]_{\star}=J^b \; ,
\end{equation}
in which the NC strength field tensor is
\begin{equation}
	\mathcal{F}_M^{ab}=\partial^aA^b-\partial^bA^a-i\left[ \, A^a \, , \, A^b \, \right]_{\star} \; .
\end{equation}
We will compare along the text some results which are in agreement with the complete work done using the Moyal picture \cite{Rivelles}. We will introduce now the plane-wave solutions in the presence of an external (constant and uniform) magnetic field in which we discuss the properties of the spatial
NC as a material medium, and how these solutions affect the dispersion relations and the group velocities.

\section{The dispersion relations and the group velocities in an external magnetic field}
\label{sec4}
Using the definition of the gauge covariant derivative (\ref{a7}) in the field equation (\ref{FieldEq}), and the Lorenz gauge condition ($\partial_aA^{a}=0$), the free wave equation for the $A^{a}$-potential is given by
\begin{equation}\label{EqwaveSym}
	\Box A_b+\theta^{cd}\left(2\,\partial_cA_a\partial_d\partial^aA_b-\,\partial_cA^a\partial_d\partial_bA_a\right)+{\cal O}(\theta^2)=0
	\; ,
\end{equation}
which, for simplicity, we will consider the leading corrections in $\theta^{ab}$. The idea here is to find the wave equation in the momentum space, when one substitutes the solution correspondingly to a single harmonic PW summed to a constant MBF, namely,
\begin{equation}\label{solution}
	A_a(x)=\tilde{A}_{a} \, \sin(k\cdot x)-\frac{1}{2} \, B_{ab} \, x^b \; ,
\end{equation}
where $\tilde{A}_{a}$ is a constant and uniform amplitude of the potential, the sinus is a function of the scalar product
$k\cdot x=k_{a}x^a=\omega \, t - {\bf k}\cdot{\bf x}$, with $k^2=\omega^2-{\bf k}^2$, and $B_{ab}=(0,-\epsilon_{ijk}B_{0k})$
sets a constant and uniform MBF. It is well known that the PW and the MBF satisfy, independently, the usual wave equation for the NC case, in which the $\theta^{ab}$-contribution disappears naturally. However, the PW written as the superposition (\ref{solution}) generates terms that depend on the NC parameter and on the magnetic background due to the non-linear properties of the Poisson bracket. The electromagnetic field of the solution (\ref{solution}) is $F^{ab}(x)= \tilde{F}^{ab} \, \cos(k\cdot x)+B^{ab}$, where $\tilde{F}^{ab}=k^a\,\tilde{A}^{b}-k^b\,\tilde{A}^{a}=(-\tilde{E}_{i},-\epsilon_{ijk}\tilde{B}_{k})$
are the electric and magnetic (constants and uniforms) wave amplitudes.
In this point, we will see how the NC affects the NC electrodynamics equations, that can be interpreted as a material medium. For the spatial NC, $\theta^{ab}=( \, 0 \, , \, \epsilon^{ijk}\,\widetilde{\theta}^{k} \, )$, the free Maxwell-Poisson equations are
\begin{subequations}
	\begin{eqnarray}
		&&\hspace{-1cm}\nabla\cdot{\bf E} =\widetilde{\theta}^{k} \, \epsilon^{ijk}\left[ \partial^{i}{\bf A}\cdot\partial^{j}{\bf E}
		+\partial^{i}{\bf A}\cdot\partial^{j}(\nabla V)\right]-\widetilde{\bm\theta}\cdot \left[\, \nabla(\nabla\cdot{\bf A}) \times \nabla V \, \right] \; ,
		\label{DivE}
		\\
		&&\hspace{-1cm}\nabla\cdot{\bf B}= \widetilde{\theta}^{k}\,\epsilon^{ijk}\left[\nabla\cdot\left( \partial^{i}{\bf A}\times\partial^{j}{\bf A} \right)+\partial^{i}{\bf A}\cdot\partial^{j}{\bf B}\right] \; ,
		\label{DivB}
		\\
		&&\hspace{-1cm}\nabla \times{\bf E}+\partial_t{\bf B}=\widetilde{\theta}^{k}\epsilon^{ijk}\left[\partial^j{\bf A}\times\partial^{i}{\bf E}
		-\partial^{i}V\,\partial^{j}{\bf B}
		-\nabla\times ( \partial^{i} V \, \partial^{j}{\bf A} )
		-\partial_t\left( \partial^i{\bf A}\times\partial^j{\bf A}\right)\right],
		\label{rotE}
		\\
		&&\hspace{-1cm}\nabla \times{\bf B}-\partial_t{\bf E}
		=\widetilde{\theta}^{k} \, \epsilon^{ijk}\left[ \partial^{i}V\,\partial^{j}{\bf E}-\partial^{i}{\bf A}\times\partial^{j}{\bf B}-\partial_t\,(\partial^{i}V\partial^{j}{\bf A})-\partial^l\,(\partial^{i}A^l\partial^{j}{\bf A})\right],
		\label{rotB}
	\end{eqnarray}
\end{subequations}
in which the index $i,j,k,l=\left\{1,2,3\right\}$. We substitute the solution
(\ref{solution}) in the Eqs. (\ref{DivE}), (\ref{DivB}), (\ref{rotE}) and (\ref{rotB}),
\begin{subequations}
	\begin{eqnarray}
		&&k_{i}\,\epsilon_{ij}\,\tilde{E}_{j}=
		-\frac{1}{2}\,({\bf k\times\widetilde{\bm \theta}})\cdot({\bf B}_0\times{\bf k})\,\tilde{V}\tan(k\cdot x) \; , \;\;\;\;\;\;\;\;\;\;\;
		\label{kE}\\
		&&{\bf k}\cdot\tilde{\bf B}=\frac{1}{2}\left[(\widetilde{\bm \theta}\cdot\tilde{\bf B})- \tilde{\bf A}\cdot({\bf k}\times\widetilde{\bm \theta}) \tan(k\cdot x)\right]({\bf k}\cdot{\bf B}_{0}) \; ,
		\label{kB}\\
		&&\tilde{{\bf B}}=\frac{1}{\omega}\,({\bf k}\times\tilde{{\bf E}})+\frac{1}{2\omega}\left[(\widetilde{\bm\theta}\times\tilde{{\bf E}})({\bf k}\cdot{\bf B}_0)-({\bf k}\times\tilde{{\bf E}})(\widetilde{{\bm \theta}}\cdot{\bf B}_0)\right]\left[1+\tan(k\cdot x)\right] \; ,
		\label{BkE}
		\\
		&&\tilde{E}_i=-\omega^{-1} \, \epsilon_{ijk} \, k_{l} \, (\mu^{-1})_{jl} \, \tilde{B}_{k}
		\; ,\label{EP}
	\end{eqnarray}
\end{subequations}
where $\epsilon_{ij}$ is the electric permittivity, and $(\mu^{-1})_{ij}$ is the inverse of magnetic permeability tensor.
Inverting the permeability tensor (\ref{EP}), we obtain
\begin{subequations}
	\begin{equation}
		\epsilon_{ij}= \delta_{ij}+\frac{1}{2} \, B_{0i} \, \widetilde{\theta}_{j} \; ,\quad\text{and}\quad
		\mu_{ij}= \delta_{ij}\left[1+\frac{1}{2}\,({\bf B}_{0}\cdot\widetilde{\bm \theta}) \right]
		-\frac{1}{2} \, \widetilde{\theta}_{i} \, B_{0j}\; .
	\end{equation}
\end{subequations}
The eigenvalues of permittivity tensor are $\lambda_{1\epsilon}=\lambda_{2\epsilon}=1$ (two degenerate values) and
$\lambda_{3\epsilon}=1+({\bf B}_{0}\cdot\widetilde{{\bm \theta}})/2$. For the permeability tensor, we obtain the eigenvalues $\lambda_{1\mu}=1$ and $\lambda_{2\mu}=\lambda_{3\mu}=1+({\bf B}_{0}\cdot\widetilde{{\bm \theta}})/2$. The vacuum limit is so recovered when $\widetilde{{\bm \theta}} \rightarrow 0$, or ${\bf B}_{0} \rightarrow 0$. Therefore, the NC space behaves like a material medium, in which the permittivity and permeability both are positive for leading order in $\widetilde{\theta}$-parameter.
Since $\widetilde{{\bm \theta}} \neq 0$, the properties of wave propagation are such that the wave amplitudes
are not, in general, perpendiculars to the propagation direction $\hat{{\bf k}}$. Furthermore, the relations (\ref{kE}), (\ref{kB})
and (\ref{BkE}) depend on the space-time coordinates through the tangent function. The time-average of these relations are
\begin{subequations}
	\begin{eqnarray}
		&&\langle\tilde{{\bf B}}\rangle=\frac{1}{2\omega}\left[({\bf k}\times\tilde{{\bf E}})\left(2-\widetilde{{\bm \theta}}\cdot{\bf B}_0\right)+(\widetilde{{\bm \theta}}\times\tilde{{\bf E}})({\bf k}\cdot{\bf B}_0)\right],
		\label{BkEmed}
\\
		&&\langle{\bf k}\cdot\tilde{{\bf E}}\rangle=-\frac{1}{2}(\widetilde{{\bm \theta}}\cdot\tilde{{\bf E}})({\bf k}\cdot{\bf B}_0)
		\; ,
		\label{kEmed}
\\
		&&\langle{\bf k}\cdot\tilde{{\bf B}}\rangle=\frac{1}{2} (\widetilde{{\bm \theta}}\cdot\tilde{{\bf B}})({\bf k}\cdot{\bf B}_{0})
		\; ,
		\label{kBmed}
	\end{eqnarray}
\end{subequations}
where $\langle{\bf k}\cdot\tilde{{\bf E}}\rangle =\langle{\bf k}\cdot\tilde{{\bf B}}\rangle=0$,
when ${\bf k}$ is perpendicular to the MBF.
Considering the particular situation of $\widetilde{{\bm \theta}}=\ell^{2}\,\hat{{\bf k}}$, where $\ell$ is a length scale,
the NC contribution is null in the relation (\ref{BkEmed}), and if ${\bf k}$ is perpendicular to ${\bf B}_{0}$
(or the external ${\bf B}_0$ field is turned off), the divergences of electric and magnetic amplitudes are
like the commutative case.
Back to PW solution, the superposition (\ref{solution}) yields again the wave equation $k^2\tilde{A}_{a}=0$,
like in the Maxwell's electrodynamics. However, for the Maxwell-Poisson equations, an interference term emerges
due to the NC contribution and the MBF. Substituting the solution (\ref{solution}) in the wave equation (\ref{EqwaveSym}), we obtain $M_{ab}\tilde{A}^b=0$,
where the wave matrix $M_{ab}$ in the momentum space is
\begin{equation}\label{EqwaveAbackground}
	M_{ab}=\left(-k_c\,k^c+\theta^{cd} \, B_{ec} \, k^e \, k_d\right)\eta_{ab}
	-\frac{1}{4} \, \theta^{cd} \, B_{bc} \, k_d \, k_a \; .
\end{equation}
The DRs are calculated by the non-trivial solution of $M_{ab}\tilde{A}^b=0$. For spatial NC,
the determinant of $M_{ab}$ is
\begin{equation}\label{detMB}
	\det(M_{ab})=(\omega^2-{\bf k}^2)^3\,[ \, {\bf k}^2-\omega^2+ \frac{15}{4} \left({\bf B}_0\times{\bf k}\right)\cdot(\widetilde{{\bm \theta}}\times{\bf k})\,] \; .
\end{equation}
The non-trivial solutions of the wave equation impose that $\det(M_{ab})=0$, where the result
(\ref{detMB}) yields the usual photon DR, namely, $\omega_{1}({\bf k})=|{\bf k}|$, and
the DR modified by the spatial NC parameter
\begin{equation}\label{omega2B}
	\omega_{2}({\bf k})=
	|{\bf k}| \, \sqrt{1+\frac{15}{4} \, ({\bf B}_0\times\hat{{\bf k}})\cdot(\widetilde{{\bm \theta}}\times\hat{{\bf k}})}\simeq|{\bf k}| \left[ 1+\frac{15}{8}\,({\bf B}_0\times\hat{{\bf k}})\cdot(\widetilde{{\bm \theta}}\times\hat{{\bf k}}) \right] \, ,
\end{equation}
where we have used a small $\widetilde{\theta}$-parameter.
The correspondent group velocity is
\begin{equation}\label{Vg}
	{\bf V}_g=\frac{{\bf k}}{\omega}\left[1
	+\frac{\frac{15}{2}({\bf k}^2-\omega^2)({\bf B}_0\cdot\widetilde{{\bm \theta}})}{8\,({\bf k}^2-\omega^2) +\frac{45}{2}\,({\bf B}_0\times{\bf k})\cdot(\widetilde{{\bm \theta}}\times{\bf k})}\right]
	+\frac{15}{4}\left[ \frac{ (\omega^2-{\bf k}^2) [ \, ({\bf B}_{0}\cdot{\bf k}) \, \widetilde{{\bm \theta}} + (\widetilde{{\bm \theta}}\cdot{\bf k}) \, {\bf B}_{0} \, ] }{8\,({\bf k}^2-\omega^2) +\frac{45}{2}\,({\bf B}_0\times{\bf k})\cdot(\widetilde{{\bm \theta}}\times{\bf k})}\right] \; .
\end{equation}
Evaluating (\ref{Vg}) in the DRs $\omega_1=|{\bf k}|$ and $\omega_2$ of (\ref{omega2B}), the first case is the same result
from usual ED, $\left. {\bf V}_g \right|_{\omega_1} = \hat{{\bf k}}$. The second DR (\ref{omega2B}) yields
the group velocity
\begin{equation}\label{Vgomega2}
	\left. {\bf V}_g \right|_{\omega_2}\simeq\hat{{\bf k}}\left[1-\frac{15}{4}({\bf B}_0\times\hat{{\bf k}})\cdot(\widetilde{{\bm \theta}}\times\hat{{\bf k}})\right]
	-\frac{15}{8}\left[({\bf B}_{0}\cdot\hat{{\bf k}}) \, \widetilde{{\bm \theta}}
	+ (\widetilde{{\bm \theta}}\cdot\hat{{\bf k}}) \, {\bf B}_{0}\right] \, .
\end{equation}
Notice that the photon DR and the group velocity depend on the direction that the vectors ${\bf B}_{0}$ and $\widetilde{{\bm \theta}}$
have with respect to the wave propagation direction $\hat{{\bf k}}$. The usual results of Maxwell ED are recovered when the MBF is null, {\it i.e.},
${\bf B}_{0} \rightarrow 0$. In this sense, the refractive index of this medium also changes with the direction of three vectors $\widetilde{\bm \theta}$, $\hat{\bf k}$ and ${\bf B}_{0}$. Thereby, the NC space is a medium that changes the propagation direction of the group velocity \cite{Rivelles}. In the next section, we show the conservation laws  in the context of a spatial NC parameter.

\section{The energy-momentum tensor}

In this section, we investigate the conservation laws associated with the NC field equations (\ref{FieldEq}) and (\ref{Bianchi}).
The components of the EMT yield the conserved physical quantities interpreted as the energy and flux densities of NC gauge theory,
when one considers only spatial NC. Applying the operator $\mathcal{D}_a$ in both sides of (\ref{FieldEq}),
the classical source is covariantly conserved
\begin{equation}
	\mathcal{D}_aJ^a=0 \; ,
\end{equation}
since the covariant derivative commute, $\left[ \, {\cal D}_a \, , \, {\cal D}_b \, \right]=0$, which means that $\mathcal{D}_{b}\mathcal{D}_{a}\mathcal{F}^{ab}=0$. In fact, if we define the bar current
$\bar{J}^{b}=J^{b}-\{ \, A_{a} \, , \, {\cal F}^{ab} \, \}$, it is conserved as is well known in Maxwell's electrodynamics.
The canonical Poisson EMT is obtained combining the NC field equations (\ref{FieldEq}) and (\ref{Bianchi}). Contracting the Bianchi identity (\ref{Bianchi}) with the field strength (\ref{strength}),
\begin{equation}
	\mathcal{D}_a\left(\mathcal{F}^{ac}\mathcal{F}_c^{\;\;\;b}+\frac{1}{4}\,\eta^{ab}\mathcal{F}^{cd}\mathcal{F}_{cd}\right)=\mathcal{F}^{ab}J_a \; .
\end{equation}
Using the properties of the covariant derivative operator, and the source field equation (\ref{FieldEq}), we obtain the conservation law \cite{emtensor}
\begin{equation}	
	{\cal D}_a{\cal T}^{ab}=J_a\,{\cal F}^{ab} \; .
\end{equation}
In the absence of external sources, the previous equation is reduced to ${\cal D}_a{\cal T}^{ab}=0$,
where the EMT is given by
\begin{equation}
	\mathcal{T}^{ab}=\mathcal{F}^{ac}\,\mathcal{F}_{c}^{\;\,b}-\eta^{ab}\,\mathcal{L}_{ED} \; ,
	\label{canonicalenergy}
\end{equation}
where $\mathcal{L}_{ED}$ is the Lagrangian (\ref{LED}) with $J^{a}=0$. This tensor transforms covariantly
under the gauge transformation $\delta_f$, {\it i. e.},
\begin{equation}
	\delta_f\mathcal{T}^{ab}=\{\mathcal{T}^{ab}\,,\,f\} \; .
\end{equation}
It is worth to note that, for the spatial NC, the EMT provides the conserved components. Considering the expressions for the gauge covariant derivative (\ref{a7}), the covariantly conserved EMT can be written as
\begin{equation}
	{\cal D}_a{\cal T}^{ab}=\partial_a{\cal T}^{ab}+\{A_a\,,\,{\cal T}^{ab}\}=0 \; ,
\end{equation}	
in which the canonical Poisson bracket can be written as a total derivative
\begin{equation}\label{partialTab}
	\partial_a{\cal T}^{ab}+\partial_d\left[\,\theta^{cd}(\partial_cA_a){\cal T}^{ab}\,\right]=0 \; .
\end{equation}
For the spatial NC, $\theta^{ab}=( \, 0 \, , \, \epsilon^{ijk}\,\widetilde{\theta}^{k} \, )$, the right side of (\ref{partialTab})
is a divergence term. When (\ref{partialTab}) is integrated over all space, the divergence terms go to zero for well behaved functions (of Schwartz's class). Consequently, the components $\mathcal{T}^{a0}=(\mathcal{T}^{00},\mathcal{T}^{i0})$, when integrated over all space, yield
the conserved quantities
\begin{equation}\label{Pa}
	P^{a}=\int dV \, {\cal T}^{a0} \; ,
\end{equation}
where $P^{a}=(P^{0},P^{i})$ are the total energy and the total momentum of the canonical Poisson EMT in the case
of a spatial NC parameter. Thereby, the energy density is $u:=\mathcal{T}^{00}$, and $S^{i}=\mathcal{T}^{i0}$ 
are the components of the Poynting vector. Note that this is a bit different from the ref. \cite{Rivelles}, in 
which they used the Seiberg-Witten map to obtain an on-shell conserved EMT.
Explicitly, the EMT can be written in terms of the gauge field $A^{a}$ as
\begin{equation}	
	\mathcal{T}^{ab}=F^{ac}\,F_{c}^{\;\,b}+\frac{1}{4}\eta^{ab}F_{cd}F^{cd}+\epsilon^{ijk}\widetilde{\theta}^k \partial_iA_c\!\left(F^{ac}\,\partial_jA^b
	+F^{bc}\partial_jA^a
	+\frac{1}{2} \, \eta^{ab} \, F^{cd}\partial_jA_d \right)+{\cal O}(\widetilde{\theta}^2) \; ,
\end{equation}
in which the conserved components in the linear approximation of $\widetilde{\theta}^k$ are
\begin{subequations}
	\begin{eqnarray}
		&&u= \frac{1}{2}\left({\bf E}^2+{\bf B}^2\right)+ \epsilon^{ijk}\widetilde{\theta}^k \left[3\left({\bf E} \cdot \partial_i {\bf A}\right)\partial_jV
		+{\bf B}\cdot(\partial_i{\bf A}\times\partial_j{\bf A})\right] \; ,
		\label{densityu}
\\
		&&S^{i}=({\bf E}\times{\bf B})^{i}
		-\epsilon^{jkl}\widetilde{\theta}^l\left[\left({\bf E}\cdot\partial_j{\bf A}\right)\partial_kA^i
		- (\partial_j {\bf A}\times{\bf B})^i \partial_kV\right].
		\label{Si}
	\end{eqnarray}
\end{subequations}

These results show that the NC parameter corrects the energy density (even if it is a small contribution), and affects
the propagation direction of the energy flux interpreted by the Poynting vector. The $S^{i}$-expression depends on the $\widetilde{\theta}^k$
in (\ref{Si}) that can change the propagation direction of the energy flux.
The next step is to investigate the PW solutions under a MBF when it is substituted in the energy density and in the Poynting vector. After the calculus of the time average, the conserved components of the EMT are
\begin{equation}
	\langle
	u
	\rangle|_A\approx\frac{1}{4}\,\tilde{{\bf E}}^2\left[1+\frac{{\bf k}^2}{\omega^2}
	-\frac{1}{\omega^2} ({\bf k}\times{\bf B}_0)\cdot({\bf k}\times\widetilde{{\bm \theta}}) \right]+\frac{1}{4\omega}[\tilde{{\bf E}}\cdot({\bf B}_{0}\times{\bf k})][\tilde{{\bf A}}\cdot(\widetilde{{\bm \theta}}\times{\bf k})]+\frac{{\bf k}^2}{2\omega}(\tilde{{\bf A}}\cdot{\bf B}_0)(\tilde{{\bf E}}\cdot\widetilde{{\bm \theta}})\,,
\end{equation}
and
\begin{eqnarray}
	&&\hspace{-0.9cm}\langle
	{\bf S}
	\rangle|_A\approx\frac{\tilde{{\bf E}}^2}{2\omega} \, {\bf k} \, (1-\widetilde{{\bm \theta}}\cdot{\bf B}_0)
	+\frac{({\bf k}\cdot{\bf B}_0)}{4}\left(\frac{\tilde{{\bf E}}^2}{\omega}+\tilde{{\bf A}}\cdot\tilde{{\bf E}} \right)\widetilde{{\bm \theta}}\nonumber\\
	&&-\frac{1}{4} \left[ {\bf k} (\widetilde{{\bm \theta}}\cdot{\bf B}_0)(\tilde{{\bf A}}\cdot\tilde{{\bf E}})+ \tilde{V}{\bf B}_0\,({\bf k}\times\widetilde{{\bm \theta}})\cdot{\bf B}_0-\tilde{V} \, {\bf B}_0^2 \, ({\bf k}\times\widetilde{{\bm \theta}})+\tilde{{\bf A}} \, ({\bf B}_0\times\tilde{{\bf E}})\cdot({\bf k}\times\widetilde{{\bm \theta}})\right].
\end{eqnarray}
We will analyze some cases of interest when the vectors
${\bf k}$, $\widetilde{{\bm \theta}}$ and ${\bf B}_{0}$ are parallel or perpendicular among themselves.
\begin{enumerate}
	\item When ${\bf k}$, $\widetilde{{\bm \theta}}$ and ${\bf B}_{0}$ are parallel, the energy density and the Poynting vector are given by
	\begin{equation}
		\langle
		u
		\rangle|_A\approx\frac{1}{4}\,\tilde{{\bf E}}^2\left(1+\frac{{\bf k}^2}{\omega^2}
		\right)+\frac{{\bf k}^2}{2\omega}(\tilde{{\bf A}}\cdot{\bf B}_0)(\tilde{{\bf E}}\cdot\widetilde{{\bm \theta}}) \; ,
	\end{equation}
	and
	\begin{equation}
		\langle
		{\bf S}
		\rangle|_A\approx\frac{\tilde{{\bf E}}^2}{2\omega} \,  (1-\widetilde{ \theta}B_0){\bf k} \,
		+\frac{\tilde{{\bf E}}^2}{4\omega} \, (k B_0) \, \widetilde{{\bm \theta}}-\frac{1}{4}\,B_{0}(\tilde{{\bf A}}\cdot\tilde{{\bf E}}) \left( \, \widetilde{\theta}\,{\bf k} \,
		-k\,\widetilde{\bm \theta} \, \right)\; ,
	\end{equation}
	where we have used a small $\widetilde{\theta}$-parameter. It is interesting to note that in the case of
	$\widetilde{{\bm \theta}}=\ell^{2}\,\hat{{\bf k}}$ mentioned previously, the contribution of $\tilde{{\bf A}}$-amplitude
	is null in the Poynting vector
	\begin{equation}
		\langle {\bf S}
		\rangle|_A \approx\frac{\tilde{{\bf E}}^2}{2\omega} \, {\bf k}
		\left( \, 1- \frac{\ell^2B_0}{2}\right) \; ,
	\end{equation}
	and the energy density, when we rescale the terms, is positive definite for a small NC
	\begin{equation}
		\langle u \rangle|_A\approx\frac{1}{4}\,\tilde{{\bf E}}^2\left(1+\frac{{\bf k}^2}{\omega^2}
		\right)
		+\frac{{\bf k}^2}{2\omega}(\tilde{{\bf A}}\cdot{\bf B}_0)(\tilde{{\bf E}}\cdot\widetilde{{\bm \theta}}) \; .
	\end{equation}
	We substitute now the DRs $\omega_1({\bf k})=|{\bf k}|$, and $\omega_2({\bf k})$ given by Eq. (\ref{omega2B}).
	The results are the same for both frequencies, since the contribution of $\widetilde{\theta}$ is negligible for $\omega_{2}({\bf k})$, we have the expressions
	\begin{equation}
		\langle u \rangle|_A\approx\frac{1}{2}\,\tilde{{\bf E}}^2
		+\frac{k}{2}(\tilde{{\bf A}}\cdot{\bf B}_0)(\tilde{{\bf E}}\cdot\widetilde{{\bm \theta}}) \; ,
\quad\text{and}\quad
		\langle{\bf S}\rangle|_A\approx\frac{\tilde{{\bf E}}^2}{2} \, \hat{{\bf k}}
		\left( \, 1- \frac{\ell^2B_0}{2}\right)\; .
	\end{equation}
	\item The case of  ${\bf k}$, ${\bf B}_{0}$ and $\widetilde{{\bm \theta}}$ perpendicular among themselves yields the results
	\begin{equation}\label{ukBperp}
		\langle u\rangle|_A\approx\frac{1}{4}\,\tilde{{\bf E}}^2\left(1+\frac{{\bf k}^2}{\omega^2}
		\right)
		+\frac{{\bf k}^2}{2\omega}(\tilde{{\bf A}}\cdot{\bf B}_0)(\tilde{{\bf E}}\cdot\widetilde{{\bm \theta}})+\frac{1}{4\omega}[\tilde{{\bf E}}\cdot({\bf B}_{0}\times{\bf k})][\tilde{{\bf A}}\cdot(\widetilde{{\bm \theta}}\times{\bf k})] \; ,
	\end{equation}
	and
	\begin{equation}
		\langle{\bf S}\rangle|_A\approx\frac{\tilde{{\bf E}}^2}{2\omega} \, {\bf k}
		+\frac{\tilde{V}}{4} \left[ \, {\bf B}_0 \cdot(\widetilde{{\bm \theta}}\times{\bf k}){\bf B}_0- {\bf B}_0^2 \, (\widetilde{{\bm \theta}}\times{\bf k}) \, \right]\; .
	\end{equation}
	It is interesting to observe that, in this case, the energy flux has no correction in the NC parameter, and the Poynting vector reduces to the usual result of the Maxwell's ED
	\begin{equation}
		\langle{\bf S}\rangle|_A \approx\frac{\tilde{{\bf E}}^2}{2\omega} \, {\bf k} \; ,
	\end{equation}
	while the energy density (\ref{ukBperp}) is positive definite. Substituting the DRs $\omega_1({\bf k})=|{\bf k}|$, and $\omega_2({\bf k})$ from (\ref{omega2B}), the results are the same for both frequencies,
	\begin{equation}
		\langle u \rangle|_{A,\omega_1}=\langle u \rangle|_{A,\omega_2}\approx\frac{1}{2}\,\tilde{{\bf E}}^2
		+\frac{k}{2}(\tilde{{\bf A}}\cdot{\bf B}_0)(\tilde{{\bf E}}\cdot\widetilde{{\bm \theta}})+\frac{1}{4}\,[\tilde{{\bf E}}\cdot({\bf B}_{0}\times\hat{{\bf k}})][\tilde{{\bf A}}\cdot(\widetilde{{\bm \theta}}\times{\bf k})] \; ,
	\end{equation}
	and
	\begin{equation}
		\langle {\bf S}\rangle|_{A,\omega_1}=\langle {\bf S}\rangle|_{A,\omega_2} \approx\frac{\tilde{{\bf E}}^2}{2} \, \hat{{\bf k}}\; .
	\end{equation}
	\item The third case of interest is when the MBF is parallel to the NC parameter, but it remains perpendicular to
	the wave propagation direction. Under these conditions, we obtain
	\begin{equation}\label{ukB0para}
		\langle u\rangle|_A \approx \frac{1}{4}\,\tilde{{\bf E}}^2\left[1+\frac{{\bf k}^2}{\omega^2}\left(1-\tilde{{\theta}}B_0\right)\right]+\frac{{\bf k}^2}{2\omega}(\tilde{{\bf A}}\cdot{\bf B}_0)(\tilde{{\bf E}}\cdot\widetilde{\bm\theta})+\frac{1}{4\omega}[\tilde{{\bf E}}\cdot({\bf B}_{0}\times{\bf k})][\tilde{{\bf A}}\cdot(\widetilde{{\bm \theta}}\times{\bf k})]\; ,
	\end{equation}
	and
	\begin{equation}
	\langle{\bf S}\rangle|_A \approx \frac{\tilde{{\bf E}}^2}{2\omega} \, \,(1-\tilde{{\theta}}B_0){\bf k}+\frac{1}{4}B_{0}\left[\tilde{V}B_0\,({\bf k}\times\widetilde{{\bm \theta}})-\tilde{\theta} \, \tilde{{\bf E}} \times ({\bf k} \times \tilde{{\bf A}})\right] \; .
	\end{equation}
	We observe that, in this case, the energy flow has a correction on the NC parameter, which can be controllable by the MBF.
	If the direction of ${\bf B}_0$ is approximate to the wave propagation one, the energy density (\ref{ukB0para})
	is positive definite for a small NC, after the rescale of expression is done.
	Substituting the DRs $\omega_1({\bf k})$ and $\omega_2({\bf k})$, we will obtain different results for each frequency, since the NC contribution is not null for the second wave frequency. This fact will play a fundamental role in the next section, resulting on the birefringence phenomenon.
\end{enumerate}
\section{The birefringence phenomenon in canonical PGT}
In this section, we investigate the possible birefringence phenomenon that emerges from the canonical PGT.
We discuss the non-linear contribution of the NC space-time, similar to the refs. \cite{Neves2,Neves3}.
The non-trivial solutions from the wave equation (\ref{EqwaveAbackground})
are given by the null determinant of the matrix $M_{ab}$, that leads to the DRs of the theory.
The results depend on the direction of vectors ${\bf k}$, ${\bf B}_{0}$ and $\widetilde{{\bm \theta}}$, in the case of
the purely spatial NC parameter. Notice that, if ${\bf k}$ is parallel to ${\bf B}_{0}$, the second DR is
$\omega_{2\parallel}({\bf k})=|{\bf k}|$. The simplest case in which the $\widetilde{\theta}$-parameter contributes
is the third in the last section. Thus, the DR (\ref{omega2B}) is reduced to
\begin{equation}
	\omega_{2\perp}({\bf k})
	\simeq |{\bf k}| \left( \, 1+\frac{15}{8}\, B_{0} \, \widetilde{\theta} \, \right) \; .
\end{equation}
We define the correspondent refractive index as $n_{\parallel}=|{\bf k}|/\omega_{2\parallel}=1$, and
$n_{\perp}=|{\bf k}|/\omega_{2\perp}$ in these situations. Therefore, the birefringence of ${\bf B}_{0}$ in relation to
${\bf k}$ is defined by the difference
\begin{equation}
	\Delta n_{B_{0}}=n_{\parallel}-n_{\perp}\simeq \frac{15}{8} \, B_0 \, \widetilde{\theta} \; .
\end{equation}
We highlight that these results are completely new, since that Mariz T. et al did not find any birefringence phenomenon on their work \cite{Rivelles}.
Using the bound from polarization vacuum with laser (PVLAS) experiment $\Delta n_{B}/B_{ext}^2= \left( \, 19 \, \pm \, 27\, \right) \times 10^{-24} \, \mbox{T}^{-2}$ for an external magnetic field of $B_{ext}=2.5 \, \mbox{T}$ \cite{25years}, we obtain the result
\begin{equation}
	\sqrt{ \widetilde{\theta} } \simeq 1.92 \times (10 \, \mbox{TeV})^{-1} \; .
\end{equation}
This result is consistent with the bounds on the $\theta$-parameter in the NC QED discussed in the ref. \cite{Carroll}.
\section{Conclusions}
We investigate the properties of the NC eletrodynamics in the canonical PGT approach. As reported in literature, PGT is the semi-classical limit of the full NC gauge theory, where the NC parameter depends on the space-time coordinates \cite{Kup33}. Thereby, the formulation of the canonical version defines a closed gauge algebra, whose the correspondent Lagrangian respects the gauge symmetry principle, and the dynamical equations lead to conservation laws in which physical quantities are affected by the NC space-time. Thus, the analysis of the classical solutions, as the PW propagation in a MBF, motivates us to understand the full NC picture \cite{Rivelles}. In this sense, the goal of this paper is the development of the field equations and the conservation laws as a first step in the construction of a full NC electrodynamics \cite{PhysRep}.
Under the PW solutions, we obtain some properties of the wave propagation, as the DRs, and the wave group velocity for a spatial NC medium. The group velocity vector has contributions of the $\widetilde{\theta}$-direction, and also of the MBF in wave propagation. The NC space behaves like a material medium in which the permittivity and permeability tensors yield characteristics that depend on the NC parameter, and on the MBF \cite{Rivelles}. When the NC parameter, or the MBF, are turned off, all these new effects go to zero, and the known results of the Maxwell's electrodynamics are recovered.
At the end of this letter, we substitute the PW solution summed to the MBF in the conserved components of the EMT. We calculate the spatial NC contribution for the energy density and for the Poynting vector in terms of the three vectors: ${\bf k}$ (wave propagation direction), ${\bf B_{0}}$ (magnetic background field) and $\widetilde{\bm \theta}$ (vector for a spatial NC) that consequently change the direction of the energy flux density for the PW. Then, we examine briefly a birefringence phenomenon associated with the directions of ${\bf k}$ and ${\bf B}_0$. Using the known result from PVLAS experiment for birefringence \cite{25years}, we obtain an estimation of $\sqrt{ \widetilde{\theta} } \simeq 1.92 \times (10 \, \mbox{TeV})^{-1}$ for the spatial NC parameter.
The perspective for forthcoming projects is to investigate the semi-classical limit of the NC gauge theories in which the $\Theta^{ab}$-parameter is a function of the space-time coordinates \cite{Kupriyanov:2020sgx}-\cite{Kup25}. There are some examples already treated on the PGT that can be consistent with the NC electrodynamics, like the $\kappa$-Minkowski space-time \cite{KKV,Kup24,Kup25}, which may be used for upcoming investigations. Moreover, it would be worth to achieve models of non-linear electrodynamics, like Born-Infeld and Mod-Max \cite{Neves2}-\cite{Neves3}, to obtain a relation with the non-linearity that emerges on the Maxwell-Poisson equations.
\section*{Acknowledgements}
We are grateful to Vladislav Kupriyanov for early discussions on this subject and the valuable remarks.
This study was partially financed by the Coordena\c{c}\~ao de Aperfei\c{c}oamento de Pessoal de N\'ivel Superior (CAPES, Brazil) - Finance Code 001.

\end{document}